\documentclass[aps,prl,reprint,amsmath,amssymb,superscriptaddress]{revtex4-2}
\usepackage{overpic}
\usepackage{graphicx}
\usepackage{dcolumn}
\usepackage{bm}
\usepackage{color}
\usepackage{float}

\begin{document}

\preprint{APS/123-QED}

\title{Achieving High Efficiency And Enhanced Beam Quality In Laser Wakefield Acceleration}

\author{Jia Wang}
\affiliation{Institute of High Energy Physics, Chinese Academy of Sciences, Beijing 100049, China}
\affiliation{State Key Laboratory of High Field Laser Physics and CAS Center for Excellence in Ultra-intense Laser Science, Shanghai Institute of Optics and Fine Mechanics (SIOM), Chinese Academy of Sciences (CAS), Shanghai 201800, China}

\author{Ming Zeng}
\email[Corresponding author:]{zengming@ihep.ac.cn}
\affiliation{Institute of High Energy Physics, Chinese Academy of Sciences, Beijing 100049, China}

\author{Dazhang Li}
\email[Corresponding author:]{lidz@ihep.ac.cn}
\affiliation{Institute of High Energy Physics, Chinese Academy of Sciences, Beijing 100049, China}
 
\author{Wentao Wang}
\email[Corresponding author:]{wwt1980@siom.ac.cn}
\affiliation{State Key Laboratory of High Field Laser Physics and CAS Center for Excellence in Ultra-intense Laser Science, Shanghai Institute of Optics and Fine Mechanics (SIOM), Chinese Academy of Sciences (CAS), Shanghai 201800, China}

\author{Song Li}
\affiliation{State Key Laboratory of High Field Laser Physics and CAS Center for Excellence in Ultra-intense Laser Science, Shanghai Institute of Optics and Fine Mechanics (SIOM), Chinese Academy of Sciences (CAS), Shanghai 201800, China}

\author{Ke Feng}
\affiliation{State Key Laboratory of High Field Laser Physics and CAS Center for Excellence in Ultra-intense Laser Science, Shanghai Institute of Optics and Fine Mechanics (SIOM), Chinese Academy of Sciences (CAS), Shanghai 201800, China}

\author{Jie Gao}
\affiliation{Institute of High Energy Physics, Chinese Academy of Sciences, Beijing 100049, China}

\date{\today}

\begin{abstract}

Laser wakefield acceleration, characterized by the extremely high electric field gradient exceeding $100\ \rm GV/m$, is regarded as a compact and cost affordable technology for the next generation of particle colliders and light sources. However, it has always been a major challenge to effectively increase the energy transfer efficiency from the laser to the accelerated beam, while ensuring the beam quality remains suitable for practical applications.
This study demonstrates that the laser with shorter pulse duration allows for a two-step dechirping process of the accelerated electron beam with charge of nanocoulomb level. The electron beams with an energy spread of 1\% can be generated with the energy transfer efficiency of 10\% to 30\% in a large parameter space. For example, one electron beam with the energy of $420\ \rm MeV$, the charge of $5.5\rm\ nC$ and the RMS energy spread of $\sim 2\%$ can be produced using an $8.3\ \rm J$ laser pulse with $7.2\ \rm fs$ duration. 
\end{abstract}


\maketitle
Laser wakefield accelerators (LWFA)~\cite{TTajimaPRL1979} provide accelerating fields for charged particles that are orders of magnitude higher than those in conventional radio-frequency (RF) accelerators, enabling electrons to attain giga-electronvolt energies within centimeter-scale distances~\cite{GonsalvesAJPRL2019,MiaoBPRX2022}. This capability makes LWFA a promising alternative for next generation particle colliders and compact radiation sources. Over the past decades, significant progress has been achieved in improving electron beam quality~\cite{MaierPRX2020,WangJMRE2022}. 
Through techniques such as controlled injection and plasma shaping, for instance, stable beams with charges of $\sim 10\ \mathrm{pC}$ and energy spread at a few per-mille level have been demonstrated~\cite{KeLTPRL2021}.
However, the underlying physical mechanisms that enable such narrow energy spreads often inherently limit the energy transfer efficiency. In LWFAs,  the energy transfer efficiency from the laser pulse to the high quality electron beam typically lies between 0.1\% and 1\%~\cite{LeeNP2006,LeemansPRL2014,GonsalvesAJPRL2019,GotzfriedPRX2020}. This limitation arises from several physical constraints. The dependence of the wakefield structure on the laser energy implies that, even the laser energy is depleted only by half, the laser might not be able to drive the wakefield to accelerate electrons~\cite{Deckerpop1996}, preventing the electrons from being stably accelerated. The fact that the laser group velocity is lower than the speed of the electron beams causes the particles to outrun the accelerating phase, thus limiting the beam energy~\cite{LuW2007,PalastroPRL2020}. Moreover, beam loading effects can significantly reduce the accelerating gradient, particularly when accelerating high charge beams~\cite{TzoufrasPRL2008}. 

Recent efforts have sought to overcome these limitations. A scheme combining a tri-plateau plasma channel with a nonlinearly chirped laser was proposed to mitigate the laser distortion and beam dephasing, thereby enhancing transfer efficiency while using controlled dynamic beam loading to reduce correlated energy spread~\cite{ShuangRe2024}. The high energy transfer efficiencies have also been reported in the plasma telescope configuration~\cite{Zengpop2020,ZengPOP2024} and the wakefield driven by a $\rm CO_2$ laser~\cite{EnricoSR2022}. Despite these advances, achieving simultaneous improvements in both beam quality and energy transfer efficiency with a simple and controllable setup remains a critical challenge, especially for applications such as particle colliders and X-ray free-electron lasers, which demand high charge and high beam quality. 

In this Letter, we propose and demonstrate an LWFA scheme that mitigates this challenge. Our approach uniquely leverages the laser defocusing injection mechanism~\cite{WJPRAB2023} to enable efficient acceleration using an ultrashort ($\sim10\ \rm fs$) drive laser. Within this injection scheme, the reduced pulse duration is naturally combined with a smaller focal spot, sustaining a high laser amplitude and a corresponding $\rm TV/m$-scale wakefield~\cite{EsareyRMP2009} at multi-terawatt power levels. This strong wakefield suppresses beam loading for $\rm nC$-scale charges while the controlled pump depletion of the short driver induces a two-step dechirping dynamics that reduces correlated energy spread before the substantial beam dephasing occurs~\cite{WangJMRE2022}.
The optimal operating regime, with laser pulse duration as an essential control parameter, is efficiently identified through multi-objective Bayesian optimization. We demonstrate through particle in cell simulations, the simultaneous generation of high charge ($\sim\rm nC$) electron beams with $\sim 20\%$ energy transfer efficiency and $\sim 1\%$ energy spread, representing a significant step toward practical, high performance plasma accelerators.

In the concept of laser defocusing injection, the amount of injected charge is given by
\begin{equation} \label{eq:Q}
  \begin{aligned}
    Q=C\times L_{inj}n_pa_0,
  \end{aligned}  
\end{equation}
where $C=1.19\times 10^{-18}\ \rm nC\ cm^2$ is an empirical constant, $L_{inj}$ is the injection length, $n_p$ is the plasma density and the laser strength parameter $a_0$ is the peak amplitude of the normalized vector potential of the laser field $\bm{a}=e\bm{A}/m_ec^2$, with $\bm{A}$ being the amplitude of the vector potential. And $e$ is elementary charge, $m_e$ is the electron mass, $c$ is the speed of light. The injection length $L_{inj}=z_{Re}\sqrt{\mathrm{exp}(-1)\sqrt{2}a_0/(\Gamma^2 k_pw_0)-1}$, 
where $k_p=\omega_p/c$, $\omega_p$ is the plasma frequency, and $w_0$ is the laser waist at the focal point in vacuum. The effective Rayleigh length within the first oscillation period of the laser pulse envelop $z_{Re}=\pi w_0^2\Gamma^2/\lambda$, where $\Gamma=w_{0e}/w_0$ and $w_{0e}$ stands for the effective laser waist at the effective focal point for the laser in the plasma~\cite{WJPRAB2023} and $\lambda$ is the laser wavelength. 
We postulate the laser at the ending time of electron injection satisfies the matching condition $k_pw_m=2\sqrt{a_m}$~\cite{LuW2007}. The laser radius is $w_m=\mathrm{exp}(-1/2)(2w_0^2a_0^2/k_p^2)^{1/4}$ as describe by \textcite{WJPRAB2023} and $a_m$ represents the matched laser strength for the matched state. 
Additionally, we assume the laser depletion length $L_{etch}=\omega^2/\omega_p^2c\tau_{\rm FWHM}$ is related to the dephasing length $L_d=2/3\omega^2/\omega_p^2w_m$ by $L_{etch}=\alpha L_d$, where $\omega$ is the laser frequency, $\tau_{\rm FWHM}$ is the pulse duration in full-width-half-maximum (FWHM), and $\alpha$ is a dimensionless coefficient. 

By substituting the laser envelope $a=a_0\mathrm{exp}(-t^2/\tau^2)$ into the one-dimensional (1D) nonlinear wakefield equation~\cite[Eq.~(20)]{EsareyRMP2009}, we can obtain the normalized wakefield amplitude $E_p$.
 For the case where $0<\alpha\leq 1$, the maximum energy gain can be expressed as 
\begin{equation} \label{eq:delta e}
  \begin{aligned}
    \bar{E}=\frac{2}{3}(2-\alpha)\alpha E_p m_ec^2\frac{\omega^2}{\omega_p^2}\sqrt{a_m}.
  \end{aligned}  
\end{equation}
Subsequently the energy transfer efficiency for the electron beam with charge $Q$ and energy $\bar{E}$ is given by
\begin{equation} \label{eq:eta}
  \begin{aligned}
    \eta=\frac{Q}{e}\frac{\bar{E}}{\varepsilon_l},
  \end{aligned}  
\end{equation}
where $\eta$ is approximately related to $\eta_{\rm FWHM}$ as $\eta\sim1.27\times\eta_{\rm FWHM}$, where $\eta_{\rm FWHM}$ corresponds to the energy transfer efficiency from the laser to the electrons with the charge of $Q_{\rm FWHM}$ in the FWHM range of the energy spectrum, and $Q\sim 1.27\times Q_{\rm FWHM}$. 
\begin{figure}
    \centering
    \begin{overpic}[width=0.49\textwidth]{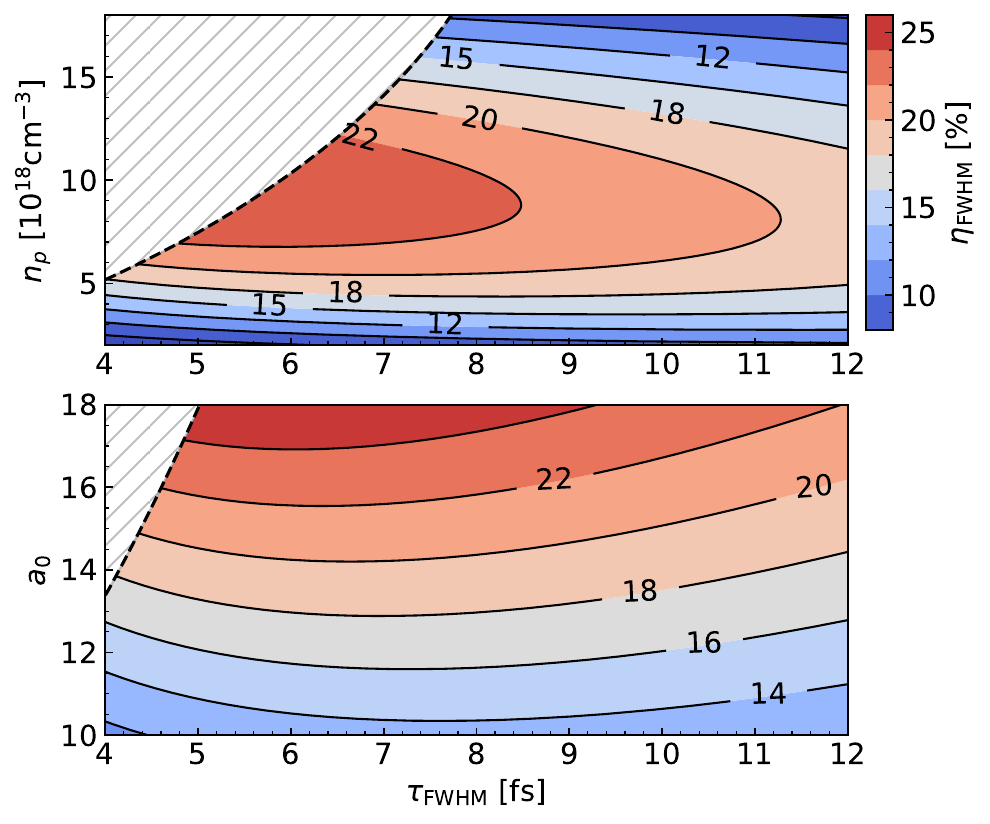}
        \put(11,78){(a)}
        \put(11,38){(b)}
    \end{overpic}
    \caption{\label{fig:eta_comp} The distribution of the energy transfer efficiency according to the model, (a) with plasma density $n_p$ and laser pulse duration $\tau_{\rm FWHM}$, while $w_0=\ \rm 6\ \mu m$ and $a_0=15$, (b) with laser strength parameter $a_0$ and laser pulse duration $\tau_{\rm FWHM}$, while $w_0=\ \rm 6\ \mu m$ and $n_p=6\times 10^{18}\ \rm cm^{-3}$. The practical parameters should be on the right side of the dashed lines $\tau_{\rm FWHM}=2 L_{inj}w_p^2/cw^2$, so that the laser pump-depletion length is sufficient for the acceleration.
}
\end{figure}
We show the theoretical energy transfer efficiency according to this model in Fig.~\ref{fig:eta_comp}. For the laser and plasma parameters of $a_0=15$, $w_0=6\ \rm \mu m$ and $n_p=6\times10^{18}\rm cm^{-3}$, the Eq.~(\ref{eq:eta}) indicates that the energy transfer efficiency attains its maximum value $\eta_{\rm FWHM}\sim 21.5\%$ at a pulse duration $\tau_{\rm FWHM}\sim 6.3\ \rm fs$. This model predicts that if the pulse duration of the laser pulse can be compressed from $30\ \rm fs$ to around $10\ \rm fs$, it will be beneficial for the energy transfer efficiency of LWFA. 
Note that the laser pulse should have sufficient energy to drive the wakefield for removing the chirp of the injected beam, which means that $L_{etch}$ should be sufficiently larger than $L_{inj}$. We assert the criteria to be $L_{etch}>2L_{inj}$, or 
\begin{equation} \label{eq:lim}
  \begin{aligned}
    \tau_{\rm FWHM}> 2\frac{L_{inj}\omega_p^2}{c\omega^2}.
  \end{aligned}  
\end{equation}
Thus, the practical parameters should be on the right side of the curve $\tau_{\rm FWHM}=2 L_{inj}\omega_p^2/c\omega^2$ in Fig.~\ref{fig:eta_comp}. 

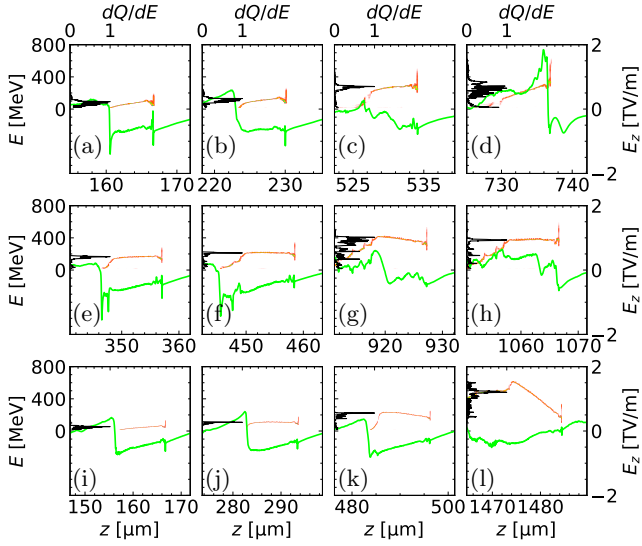
\begin{figure}[h]
    \centering
    \begin{overpic}[width=0.48\textwidth]{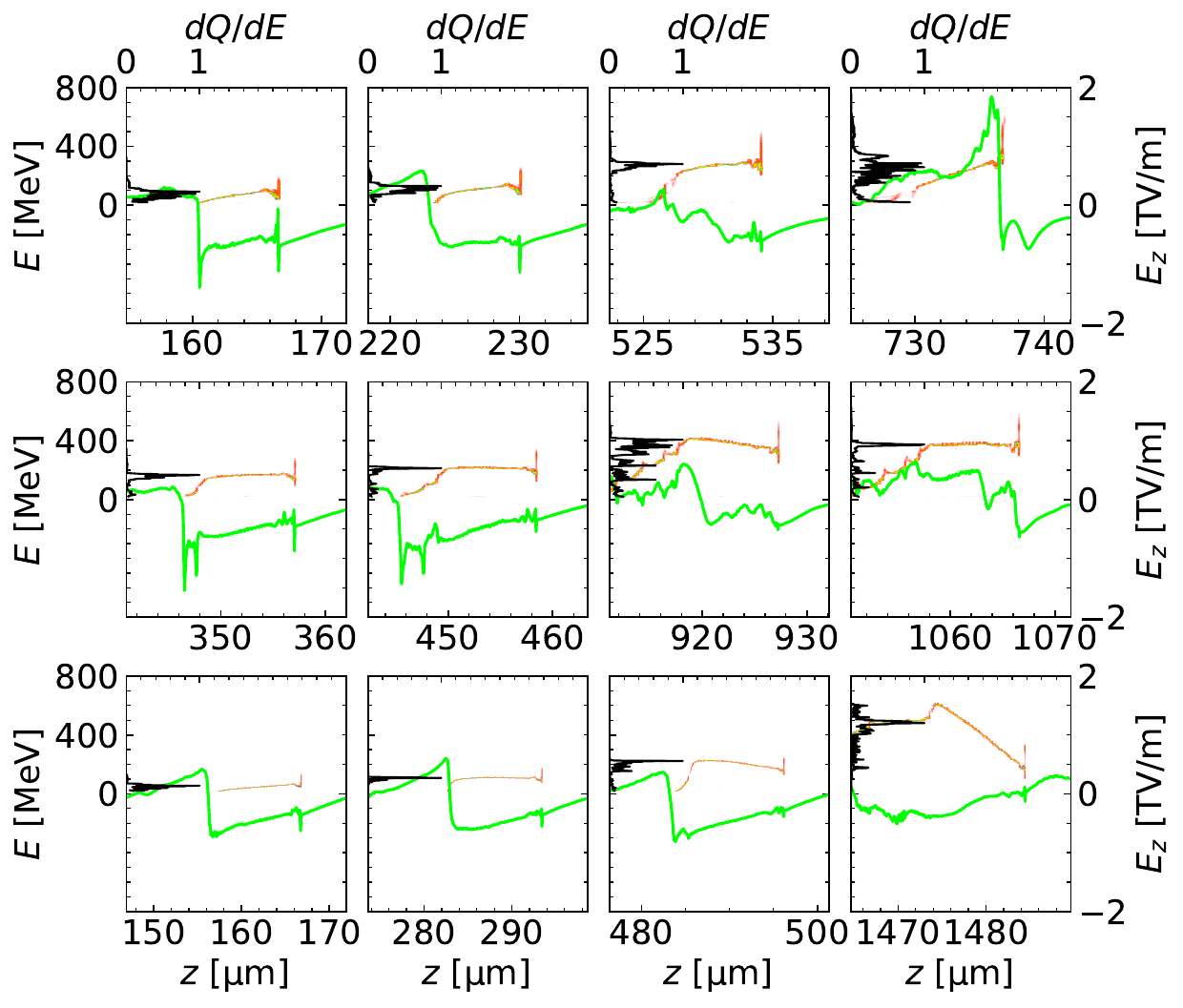}
         \put(11,61){(a)}
         \put(31.5,61){(b)}
         \put(52,61){(c)}
         \put(72.5,61){(d)}
         \put(11,35){(e)}
         \put(31.5,35){(f)}
         \put(52,35){(g)}
         \put(72.5,35){(h)}
         \put(11,9.5){(i)}
         \put(31.5,9.5){(j)}
         \put(52,9.5){(k)}
         \put(72.5,9.5){(l)}
    \end{overpic}
    \caption{\label{fig:snap_phase} 
    The phase-space distributions and the spectra (black curves) of the electron beam during the acceleration process for different laser duration cases. The distributions of longitudinal electric field $E_z$ are also shown as the green curves. The laser duration are (a)-(d) $\tau_{\rm FWHM}=5\ \rm fs$, (e)-(h) $\tau_{\rm FWHM}=8\ \rm fs$ and (i)-(l) $\tau_{\rm FWHM}=15\ \rm fs$.
}
\end{figure}

Here, we performed three quasi-3D Particle-in-Cell (PIC) simulations using the code of FBPIC~\cite{LEHEcpc2016} to show the impact of the laser pulse duration on the beam quality and examine the model above, as depicted in Fig.~\ref{fig:snap_phase}. The plasma, with a plateau density of $n_p=6\times10^{18}\rm cm^{-3}$, features an upramp with profile of $n=n_p\mathrm{exp}( -z^2/\sigma_z^2)$ spanning from $z=-500\ \rm \mu m$ to $z=0$, where $\sigma_z=200\ \rm \mu m$. The lasers have the strength parameter of $a_0=15$, the laser waist of $w_0=6\ \rm \mu m$, the wavelength of $\lambda=800\ \rm nm$ and the focusing position of $z_f=0$. The FWHM pulse duration $\tau_{\rm FWHM}$ in the 3 cases are $5\ \rm fs$, $8\ \rm fs$ and $15\ \rm fs$, respectively. The simulations have the box size of $\left(100\ \rm \mu m, 65\ \rm \mu m\right)$ and grid numbers of $\left(256, 2048\right)$ in $r$ and $z$ directions.   
For the case of $\tau_{\rm FWHM}=5\ \rm fs$, the laser pump depletion length is too short, that there is not enough length to remove the negative energy chirp of the electron beam.  At $z\sim 530\ \rm \mu m$ as shown in Fig.~\ref{fig:snap_phase}(c), the beam has charge of $Q_{\rm FWHM}\sim0.76\ \rm nC$, the average energy of $\sim 280\ \rm MeV$ and the FWHM energy spread of $\sim 6.2\%$ and the energy transfer efficiency of $\eta_{\rm FWHM}\sim15\%$, within $\rm FWHM$ range of the energy spectrum. 
For the case of $\tau_{\rm FWHM}=8\ \rm fs$, the chirp of the electron beam is removed twice at $z\sim 450\ \rm \mu m$ and $z\sim 1060\ \rm \mu m$ as shown in Fig.~\ref{fig:snap_phase}(f) and (h), respectively.
At $z\sim 1060\ \rm \mu m$, the beam obtains the maximum energy of $\bar{E}\sim 372\ \rm MeV$, and has the charge of $Q_{\rm FWHM}\sim1.27\ \rm nC$, the energy spread of $\Delta E_{\rm FWHM} / \bar{E}\sim 3.28\%$ and the energy transfer efficiency of $\eta_{\rm FWHM}\sim20\%$. Note that our model, i.e.\ Eqs.~(\ref{eq:Q}), (\ref{eq:delta e}) and (\ref{eq:eta}),  predicts the beam parameters to be $\bar{E}=360\ \rm MeV$, $Q_{\rm FWHM}=1.34\ \rm nC$ and $\eta_{\rm FWHM}=21.1\%$, which is in good agreement with the simulation result. For this moderate laser pump depletion length, the beam phase-rotation reverses from negative chirp to positive chirp, because the electric field $E_z$ changes from having a positive longitudinal derivative ($dE_z/dz>0$) to a negative one ($dE_z/dz<0$) as the laser energy is gradually consumed in the plasma. This in turn causes the longitudinal phase space (LPS) to rotate reversely, consequently canceling the beam positive chirp. This two-step dechirping phenomenon caused by the laser depletion can also be observed in other injection schemes~\cite{WangJMRE2022}.
For the case of $\tau_{\rm FWHM}=15\ \rm fs$, the initial negative energy chirp of the beam is removed at $z\sim 290\ \rm \mu m$ (Fig.~\ref{fig:snap_phase}(j)) where the beam have the charge of $Q_{\rm FWHM}\sim0.84\ \rm nC$, the energy of $\sim 109\ \rm MeV$, the FWHM energy spread of $\sim 5.06\%$, and the energy transfer efficiency of $\eta_{\rm FWHM}\sim2.1\%$. However, the reversal of the phase-rotation does not take place in this case, because the laser is still strong enough to drive a nonlinear wakefield with $dE_z/dz>0$ even the dephasing occurs, as the pump-depletion length is long in this case. Thus, the second removing of the energy chirp does not take place, and the beam quality is not as good as the $\tau_{\rm FWHM}=8\ \rm fs$ case.
Here, we respectively refer to the mechanisms corresponding to long, moderate, and short laser pump depletion lengths (pulse duration) as the first, second, and third types of dechirping mechanisms.

\begin{figure}
    \centering
    \begin{overpic}[width=0.49\textwidth]{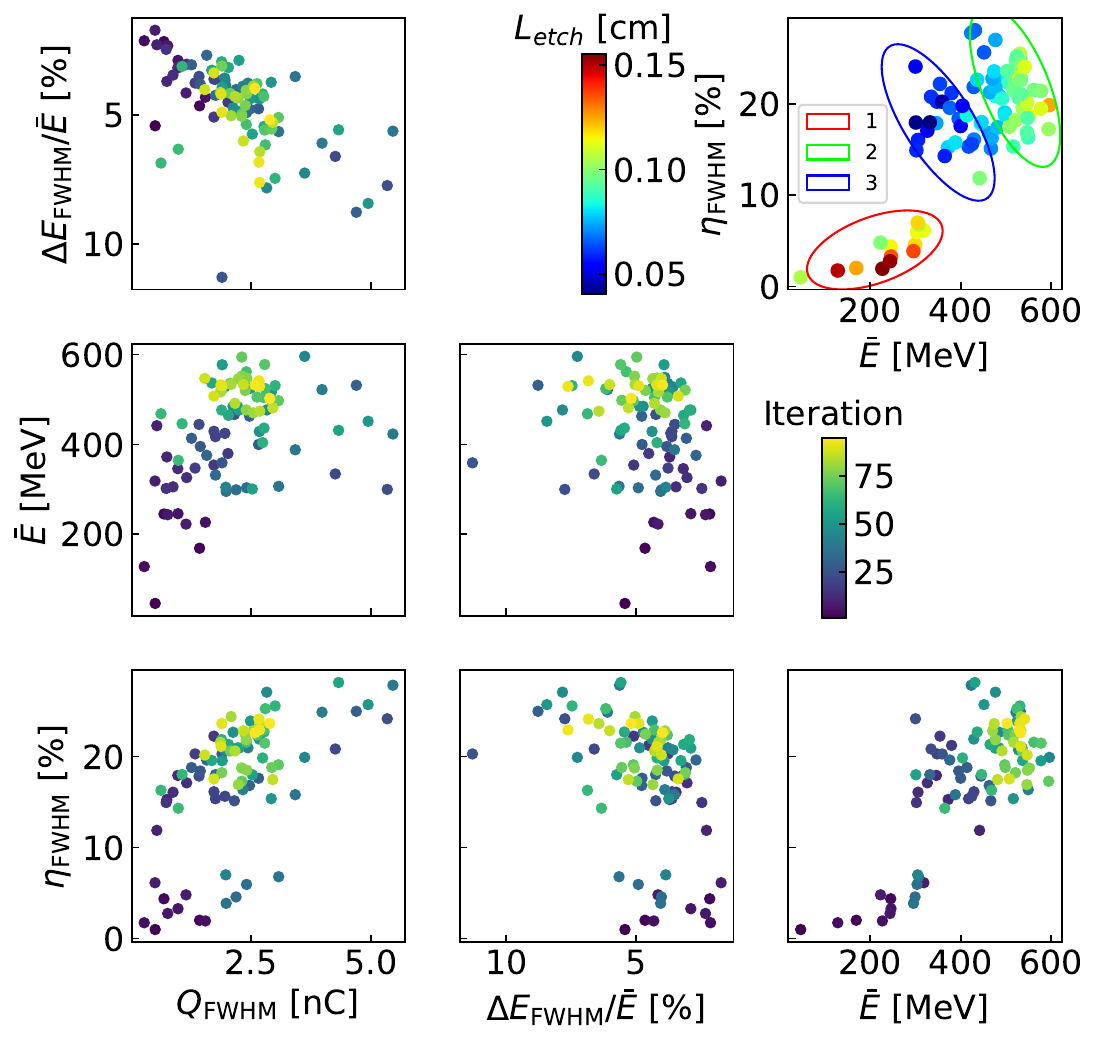}
         \put(30,71){(a)}
         \put(30,41){(b)}
         \put(42,41){(c)}
         \put(30,11){(d)}
         \put(42,11){(e)}
         \put(89,11){(f-1)}
         \put(89,71){(f-2)}
    \end{overpic}
    \caption{\label{fig:obj_vs_obj_new} Diagram of the multi-objective Bayesian optimization. (a) The objective pairs of energy spread $\Delta E_{\rm FWHM} / \bar{E}$ vs. charge $Q_{\rm FWHM}$. (b) The beam energy $\bar{E}$ vs. charge $Q_{\rm FWHM}$. (c) The beam energy $\bar{E}$ vs. energy spread $\Delta E_{\rm FWHM} / \bar{E}$. (d) The energy transfer efficiency $\eta_{\rm FWHM}$ vs. charge $Q_{\rm FWHM}$. (e) The energy transfer efficiency $\eta_{\rm FWHM}$ vs. energy spread $\Delta E_{\rm FWHM} / \bar{E}$. (f-1) The energy transfer efficiency $\eta_{\rm FWHM}$ vs. energy $\bar{E}$. In (f-2), three clusters of dots are categorized according to the associated laser depletion length $L_{etch}$ and the legend numbers are indicative of the three types of dechirping mechanisms. 
}
\end{figure}
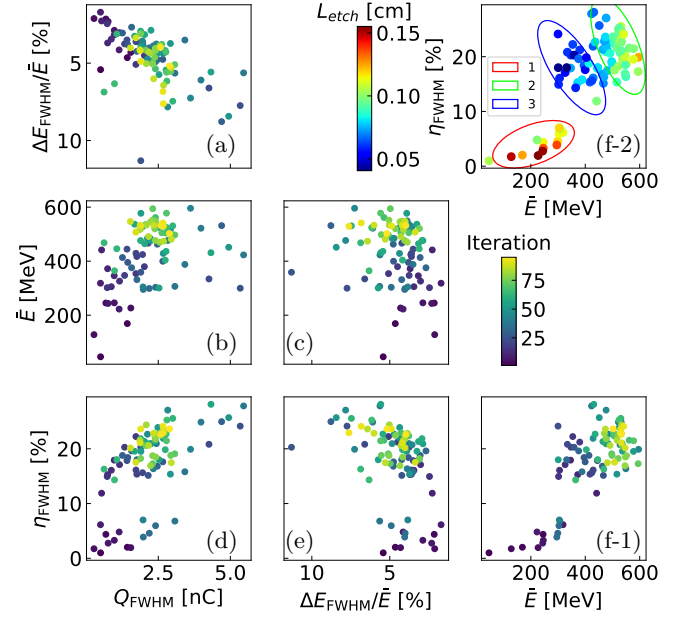 

In order to optimize the parameters of beam energy spread and energy transfer efficiency that cannot be simultaneously considered in theory, and verify the strong dependence between beam quality and laser pulse duration, multi-objective Bayesian optimization (MOBO)~\cite{IrshadPRR2023,JalasPRAB2023} is investigated for its ability to quickly evaluate and predict relied on a very small amount of data by establishing surrogate models. MOBO can balance the objective functions by finding the Pareto front of the system. 
During the optimization cycles, the plasma density $n_p$, laser strength parameter $a_0$, laser waist $w_0$ and laser pulse duration $\tau_{\rm FWHM}$ are selected as the optimized input parameters.  
The plasma density is restricted to the range of $\left[4\times 10^{18} \rm cm^{-3}, 8\times 10^{18} \rm cm^{-3}\right]$. 
Considering the currently achievable laser power in experiment and the specific requirement of triggering this injection, the varying range of laser strength $a_0$ is between 8 and 20. In the laser defocusing injection scheme, the laser waist $w_0$ should be under $10\ \rm \mu m$~\cite{WJPRAB2023}, because $w_0=10\ \rm \mu m$ is the transition from the laser defocusing injection to general self-injection. And we observed that the beam energy chirp cannot be removed if the laser waist is smaller than $4\ \rm \mu m$. So the range of laser waist $w_0$ is set to $\left[4\ \rm \mu m, 9\ \rm \mu m\right]$. 
The range of the laser pulse duration $\tau_{\rm FWHM}$ is from $5\ \rm fs$ to $20\ \rm fs$, so that all three types of dechirping mechanisms are included in the optimization iteration. It should be noted that the effective Rayleigh length $z_{Re}$ is closely related to the position of real laser focal point in plasma. The longer the interaction length between the laser pulse and the plasma before the laser reaches its real focal point, the smaller the effective Rayleigh length becomes. This, in turn, reduces the energy to which the beam can be accelerated. Therefore, for all simulations, the laser focus position is fixed at the entrance of the plasma density plateau.

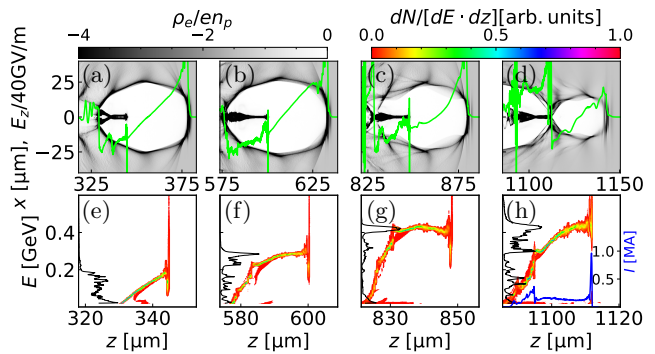
\begin{figure}
    \centering
    \vspace{0.2cm}
    \begin{overpic}[width=0.48\textwidth]{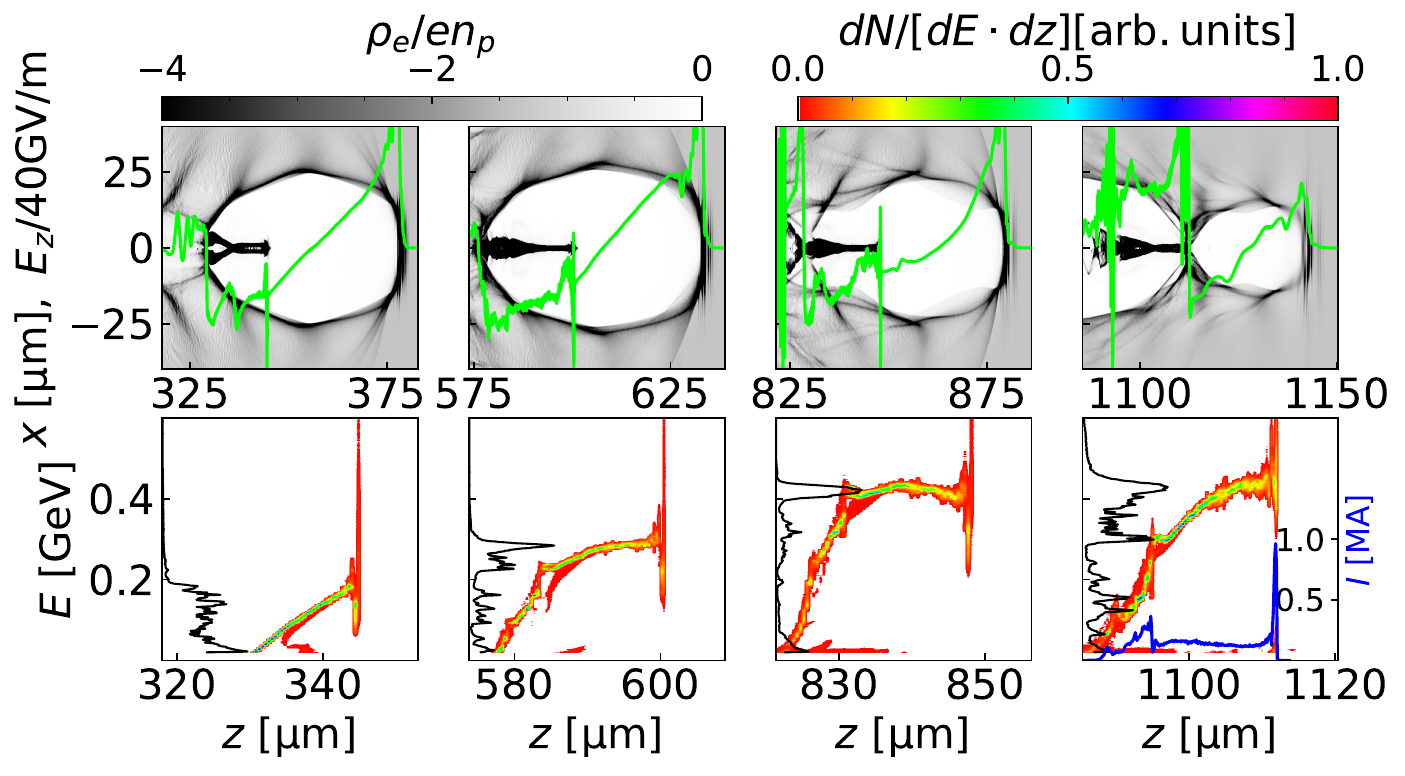}
    \put(12,43){(a)}
    \put(34,43){(b)}
    \put(56,43){(c)}
    \put(77.5,43){(d)}
    \put(12,22){(e)}
    \put(34,22){(f)}
    \put(56,22){(g)}
    \put(77.5,22){(h)}
    
    \end{overpic}
    \caption{\label{fig:lower density} The snapshots of the plasma wakefield (a)-(d) and the corresponding distribution of the LPS (e)-(h) of the injected electron beam for the simulation with plasma density $n_p=5.4\times 10^{18} \rm cm^{-3}$ and an 8.3 J laser at different positions. The line-out of the axial longitudinal electric field $E_z$ (the green curve) is in the unit of $40\ \rm GV/m$. $\rho_e$ is the charge density of the background electrons, $e$ is the elementary charge. The black curves in (e)-(h) are the energy spectrum, and the blue curve in (h) is the beam current. At $z\sim840\ \rm \mu m$, the energy transfer efficiency of $\eta_{\rm FWHM}=28\%$ is reached. And we can see the front part of the beam is in a positive chirp state, so, this situation still belongs to the second type of dechirping mechanism.
}
\end{figure}

We set four objective parameters for optimization: the charge $Q_{\rm FWHM}$, the relative energy spread $\Delta E_{\rm FWHM} / \bar{E}$, the average energy $\bar{E}$ and the corresponding energy transfer efficiency $\eta_{\rm FWHM}$. 
The beam parameters at which the chirp of the beam is approximately removed 
are chosen as the training data.  
Upon executing nearly 100 iterations of the algorithm, the objective parameters are depicted in Fig.~\ref{fig:obj_vs_obj_new}. Evidently, three separate Pareto fronts or clusters of points can be easily identified, which roughly correspond to the three types of dechirping mechanism described earlier.
The behavior of the algorithm is observed to quickly deviate from the first type of dechirping mechanism and converge to the second type of dechirping mechanism.
The elecron beams generated in the second type of dechirping mechanism display higher energy and higher energy transfer efficiency. The physical trade-off between the charge and the energy spread broadens the spectrum of the beam as the charge is much higher (Fig.~\ref{fig:obj_vs_obj_new} (a)). In the currently available results from the MOBO, the maximum energy transfer efficiency, $\eta_{\max}\sim 28\%$, is related to the second dechirping mechanism. The electron beam associated with this efficiency has $Q_{\rm FWHM}=5.5\ \rm nC$, energy spread of $\Delta E_{\rm FWHM} / \bar{E}\sim 5.6\%$, energy of $\bar{E}\sim 0.42\ \rm GeV$, average current of $\left<I\right>\sim160\ \rm kA$ and peak current of $\sim 1\ \rm MA$. This optimal result is obtained by the laser and plasma parameters of $a_0=20$, $w_0=9\ \rm \mu m$, $\tau_{\rm FWHM}=7.17\ \rm fs$ and $n_p=5.4\times 10^{18} \rm cm^{-3}$. The evolution of the bubble and the phase space of the electron beam are displayed in Fig.~\ref{fig:lower density}. This significant improvement indicates the effectiveness of the MOBO-based optimization approach in maximizing the performance of the electron beam under the laser defocusing injection mechanism.

In summary, we have performed PIC simulations to show that LWFAs driven by moderately short laser pulse duration can have enhanced beam quality and energy transfer efficiency. By applying the laser defocusing injection scheme, a sub-petawatt laser pulse with a duration around $10\ \rm fs$ generates a nanocoulomb electron beam with an initially negative chirp, which is experiencing an automatic two-step dechirping process during the subsequent acceleration process. By employing multi-objective Bayesian optimization, we show the generation of electron beams with high energy transfer efficiency of up to $28\%$, small energy spread of $\lesssim 5\%$, energy of $\sim 500\ \rm MeV$ and charge of $\sim 5\ \rm nC$ using a $10\ \rm J$ laser. Our findings make a significant step towards the compact accelerators based applications.

This work is supported by the Strategic Priority Research Program of the Chinese Academy of Sciences (Grant No.~XDB0530000, XDB0890201 and XDB0890202), the National Natural Science Foundation of China (Grant No.~12475159, 112225411, 12388102, 12474349 and  12174410), Key Research Program of Frontier Sciences of Chinese Academy of Sciences (Grant No. QYZDJ-SSW-SLH004), CAS Project for Young Scientists in Basic Research (No.~YSBR060), CAS Youth Innovation Promotion Association (No. 2022242), and the New Cornerstone Science Foundation through the XPLORER PRIZE.

\appendix

\nocite{*}

\bibliography{main}

@PREAMBLE{
 "\providecommand{\noopsort}[1]{}" 
 # "\providecommand{\singleletter}[1]{#1}%" 
}

@article{TTajimaPRL1979,
  title = {Laser Electron Accelerator},
  author = {Tajima, T. and Dawson, J. M.},
  journal = {Phys. Rev. Lett.},
  volume = {43},
  issue = {4},
  pages = {267--270},
  numpages = {0},
  year = {1979},
  month = {Jul},
  publisher = {American Physical Society},
  doi = {10.1103/PhysRevLett.43.267},
  url = {https://link.aps.org/doi/10.1103/PhysRevLett.43.267}
}

@article{SpranglePRA1990,
  title = {Nonlinear interaction of intense laser pulses in plasmas},
  author = {Sprangle, P. and Esarey, E. and Ting, A.},
  journal = {Phys. Rev. A},
  volume = {41},
  issue = {8},
  pages = {4463--4469},
  numpages = {0},
  year = {1990},
  month = {Apr},
  publisher = {American Physical Society},
  doi = {10.1103/PhysRevA.41.4463},
  url = {https://link.aps.org/doi/10.1103/PhysRevA.41.4463}
}

@article{Deckerpop1996,
    author = {Decker, C. D. and Mori, W. B. and Tzeng, K.‐C. and Katsouleas, T.},
    title = "{The evolution of ultra‐intense, short‐pulse lasers in underdense plasmas}",
    journal = {Physics of Plasmas},
    volume = {3},
    number = {5},
    pages = {2047-2056},
    year = {1996},
    month = {05},
    issn = {1070-664X},
    doi = {10.1063/1.872001},
    url = {https://doi.org/10.1063/1.872001}
}

@article{ LeeNP2006,
Author = {Leemans, W. P. and Nagler, B. and Gonsalves, A. J. and Toth, Cs. and
   Nakamura, K. and Geddes, C. G. R. and Esarey, E. and Schroeder, C. B.
   and Hooker, S. M.},
Title = {{GeV electron beams from a centimetre-scale accelerator}},
Journal = {{Nature Phys}},
Year = {{2006}},
Volume = {{2}},
Number = {{10}},
}

@article{LuW2007,
  title = {Generating multi-GeV electron bunches using single stage laser wakefield acceleration in a 3D nonlinear regime},
  author = {Lu, W. and Tzoufras, M. and Joshi, C. and Tsung, F. S. and Mori, W. B. and Vieira, J. and Fonseca, R. A. and Silva, L. O.},
  journal = {Phys. Rev. ST Accel. Beams},
  volume = {10},
  issue = {6},
  pages = {061301},
  numpages = {12},
  year = {2007},
  month = {Jun},
  publisher = {American Physical Society},
  doi = {10.1103/PhysRevSTAB.10.061301},
  url = {https://link.aps.org/doi/10.1103/PhysRevSTAB.10.061301}
}

@article{TzoufrasPRL2008,
  title = {Beam Loading in the Nonlinear Regime of Plasma-Based Acceleration},
  author = {Tzoufras, M. and Lu, W. and Tsung, F. S. and Huang, C. and Mori, W. B. and Katsouleas, T. and Vieira, J. and Fonseca, R. A. and Silva, L. O.},
  journal = {Phys. Rev. Lett.},
  volume = {101},
  issue = {14},
  pages = {145002},
  numpages = {4},
  year = {2008},
  month = {Sep},
  publisher = {American Physical Society},
  doi = {10.1103/PhysRevLett.101.145002},
  url = {https://link.aps.org/doi/10.1103/PhysRevLett.101.145002}
}

@article{EsareyRMP2009,
  title = {Physics of laser-driven plasma-based electron accelerators},
  author = {Esarey, E. and Schroeder, C. B. and Leemans, W. P.},
  journal = {Rev. Mod. Phys.},
  volume = {81},
  issue = {3},
  pages = {1229--1285},
  numpages = {0},
  year = {2009},
  month = {Aug},
  publisher = {American Physical Society},
  doi = {10.1103/RevModPhys.81.1229},
  url = {https://link.aps.org/doi/10.1103/RevModPhys.81.1229}
}

@article{McGuffeypop2012,
    author = {McGuffey, C. and Matsuoka, T. and Kneip, S. and Schumaker, W. and Dollar, F. and Zulick, C. and Chvykov, V. and Kalintchenko, G. and Yanovsky, V. and Maksimchuk, A. and Thomas, A. G. R. and Krushelnick, K. and Najmudin, Z.},
    title = "{Experimental laser wakefield acceleration scalings exceeding 100TW}",
    journal = {Physics of Plasmas},
    volume = {19},
    number = {6},
    pages = {063113},
    year = {2012},
    month = {06},
    issn = {1070-664X},
    doi = {10.1063/1.4729659},
    url = {https://doi.org/10.1063/1.4729659}
}

@article{Kalmykovpop2012,
    author = {Kalmykov, S. Y. and Beck, A. and Yi, S. A. and Khudik, V. N. and Downer, M. C. and Lefebvre, E. and Shadwick, B. A. and Umstadter, D. P.},
    title = "{Electron self-injection into an evolving plasma bubble: Quasi-monoenergetic laser-plasma acceleration in the blowout regimea)}",
    journal = {Physics of Plasmas},
    volume = {18},
    number = {5},
    pages = {056704},
    year = {2011},
    month = {04},
    issn = {1070-664X},
    doi = {10.1063/1.3566062},
    url = {https://doi.org/10.1063/1.3566062}
}

@article{LeemansPRL2014,
  title = {Multi-GeV Electron Beams from Capillary-Discharge-Guided Subpetawatt Laser Pulses in the Self-Trapping Regime},
  author = {Leemans, W. P. and Gonsalves, A. J. and Mao, H.-S. and Nakamura, K. and Benedetti, C. and Schroeder, C. B. and T\'oth, Cs. and Daniels, J. and Mittelberger, D. E. and Bulanov, S. S. and Vay, J.-L. and Geddes, C. G. R. and Esarey, E.},
  journal = {Phys. Rev. Lett.},
  volume = {113},
  issue = {24},
  pages = {245002},
  numpages = {5},
  year = {2014},
  month = {Dec},
  publisher = {American Physical Society},
  doi = {10.1103/PhysRevLett.113.245002},
  url = {https://link.aps.org/doi/10.1103/PhysRevLett.113.245002}
}

@article{LEHEcpc2016,
title = {A spectral, quasi-cylindrical and dispersion-free Particle-In-Cell algorithm},
journal = {Computer Physics Communications},
volume = {203},
pages = {66-82},
year = {2016},
issn = {0010-4655},
doi = {https://doi.org/10.1016/j.cpc.2016.02.007},
url = {https://www.sciencedirect.com/science/article/pii/S0010465516300224},
author = {Rémi Lehe and Manuel Kirchen and Igor A. Andriyash and Brendan B. Godfrey and Jean-Luc Vay}
}

@article{GonsalvesAJPRL2019,
  title = {Petawatt Laser Guiding and Electron Beam Acceleration to 8 GeV in a Laser-Heated Capillary Discharge Waveguide},
  author = {Gonsalves, A. J. and Nakamura, K. and Daniels, J. and Benedetti, C. and Pieronek, C. and de Raadt, T. C. H. and Steinke, S. and Bin, J. H. and Bulanov, S. S. and van Tilborg, J. and Geddes, C. G. R. and Schroeder, C. B. and T\'oth, Cs. and Esarey, E. and Swanson, K. and Fan-Chiang, L. and Bagdasarov, G. and Bobrova, N. and Gasilov, V. and Korn, G. and Sasorov, P. and Leemans, W. P.},
  journal = {Phys. Rev. Lett.},
  volume = {122},
  issue = {8},
  pages = {084801},
  numpages = {6},
  year = {2019},
  month = {Feb},
  publisher = {American Physical Society},
  doi = {10.1103/PhysRevLett.122.084801},
  url = {https://link.aps.org/doi/10.1103/PhysRevLett.122.084801}
}

@article{Zengpop2020,
    author = {Zeng, Ming and Martinez de la Ossa, Alberto and Poder, Kristjan and Osterhoff, Jens},
    title = "{Plasma eyepieces for petawatt class lasers}",
    journal = {Physics of Plasmas},
    volume = {27},
    number = {2},
    pages = {023109},
    year = {2020},
    month = {02},
    issn = {1070-664X},
    doi = {10.1063/1.5116416},
    url = {https://doi.org/10.1063/1.5116416}
}

@article{MaierPRX2020,
  title = {Decoding Sources of Energy Variability in a Laser-Plasma Accelerator},
  author = {Maier, Andreas R. and Delbos, Niels M. and Eichner, Timo and H\"ubner, Lars and Jalas, S\"oren and Jeppe, Laurids and Jolly, Spencer W. and Kirchen, Manuel and Leroux, Vincent and Messner, Philipp and Schnepp, Matthias and Trunk, Maximilian and Walker, Paul A. and Werle, Christian and Winkler, Paul},
  journal = {Phys. Rev. X},
  volume = {10},
  issue = {3},
  pages = {031039},
  numpages = {7},
  year = {2020},
  month = {Aug},
  publisher = {American Physical Society},
  doi = {10.1103/PhysRevX.10.031039},
  url = {https://link.aps.org/doi/10.1103/PhysRevX.10.031039}
}

@article{PalastroPRL2020,
  title = {Dephasingless Laser Wakefield Acceleration},
  author = {Palastro, J. P. and Shaw, J. L. and Franke, P. and Ramsey, D. and Simpson, T. T. and Froula, D. H.},
  journal = {Phys. Rev. Lett.},
  volume = {124},
  issue = {13},
  pages = {134802},
  numpages = {6},
  year = {2020},
  month = {Mar},
  publisher = {American Physical Society},
  doi = {10.1103/PhysRevLett.124.134802},
  url = {https://link.aps.org/doi/10.1103/PhysRevLett.124.134802}
}

@article{GotzfriedPRX2020,
  title = {Physics of High-Charge Electron Beams in Laser-Plasma Wakefields},
  author = {G\"otzfried, J. and D\"opp, A. and Gilljohann, M. F. and Foerster, F. M. and Ding, H. and Schindler, S. and Schilling, G. and Buck, A. and Veisz, L. and Karsch, S.},
  journal = {Phys. Rev. X},
  volume = {10},
  issue = {4},
  pages = {041015},
  numpages = {15},
  year = {2020},
  month = {Oct},
  publisher = {American Physical Society},
  doi = {10.1103/PhysRevX.10.041015},
  url = {https://link.aps.org/doi/10.1103/PhysRevX.10.041015}
}

@article{KeLTPRL2021,
  title = {Near-GeV Electron Beams at a Few Per-Mille Level from a Laser Wakefield Accelerator via Density-Tailored Plasma},
  author = {Ke, L. T. and Feng, K. and Wang, W. T. and Qin, Z. Y. and Yu, C. H. and Wu, Y. and Chen, Y. and Qi, R. and Zhang, Z. J. and Xu, Y. and Yang, X. J. and Leng, Y. X. and Liu, J. S. and Li, R. X. and Xu, Z. Z.},
  journal = {Phys. Rev. Lett.},
  volume = {126},
  issue = {21},
  pages = {214801},
  numpages = {6},
  year = {2021},
  month = {May},
  publisher = {American Physical Society},
  doi = {10.1103/PhysRevLett.126.214801},
  url = {https://link.aps.org/doi/10.1103/PhysRevLett.126.214801}
}

@article{WTWangNature2021,
  doi = {10.1038/s41586-021-03678-x},
  url = {https://doi.org/10.1038/s41586-021-03678-x},
  year = 2021,
  month = {jul},
  volume = {595},
  issue = {7868},
  pages = {516–520},
  author = {Wang, Wentao and Feng, Ke and Ke, Lintong and Yu, Changhai and Xu, Yi and Qi, Rong and Chen, Yu and Qin, Zhiyongv and Zhang, Zhijun and Fang, Ming and Liu, Jiaqi and Jiang, Kangnan and Wang, Hao and Wang, Cheng and Yang, Xiaojun and Wu, Fenxiang and Leng, Yuxin and Liu, Jiansheng and Li, Ruxin and Xu, Zhizhan},
  title = {Free-electron lasing at 27 nanometres based on a laser wakefield accelerator},
  journal = {Nature},
}

@article{JalasPRL2021,
  title = {Bayesian Optimization of a Laser-Plasma Accelerator},
  author = {Jalas, S\"oren and Kirchen, Manuel and Messner, Philipp and Winkler, Paul and H\"ubner, Lars and Dirkwinkel, Julian and Schnepp, Matthias and Lehe, Remi and Maier, Andreas R.},
  journal = {Phys. Rev. Lett.},
  volume = {126},
  issue = {10},
  pages = {104801},
  numpages = {6},
  year = {2021},
  month = {Mar},
  publisher = {American Physical Society},
  doi = {10.1103/PhysRevLett.126.104801},
  url = {https://link.aps.org/doi/10.1103/PhysRevLett.126.104801}
}

@article{MiaoBPRX2022,
  title = {Multi-GeV Electron Bunches from an All-Optical Laser Wakefield Accelerator},
  author = {Miao, B. and Shrock, J. E. and Feder, L. and Hollinger, R. C. and Morrison, J. and Nedbailo, R. and Picksley, A. and Song, H. and Wang, S. and Rocca, J. J. and Milchberg, H. M.},
  journal = {Phys. Rev. X},
  volume = {12},
  issue = {3},
  pages = {031038},
  numpages = {17},
  year = {2022},
  month = {Sep},
  publisher = {American Physical Society},
  doi = {10.1103/PhysRevX.12.031038},
  url = {https://link.aps.org/doi/10.1103/PhysRevX.12.031038}
}

@article{WangJMRE2022,
author = {Wang,Jia  and Zeng,Ming  and Li,Dazhang  and Wang,Xiaoning  and Lu,Wei  and Gao,Jie },
title = {Injection induced by coaxial laser interference in laser wakefield accelerators},
journal = {Matter Radiat. Extremes},
volume = {7},
number = {5},
pages = {054001},
year = {2022},
doi = {10.1063/5.0101098},

URL = { 
        https://doi.org/10.1063/5.0101098
    
},
eprint = { 
        https://doi.org/10.1063/5.0101098
    
}

}

@article{EnricoSR2022,
  doi = {10.1038/s41598-022-10160-9},
  url = {https://doi.org/10.1038/s41598-022-10160-9},
  year = 2022,
  month = {May},
  volume = {12},
  issue = {6703},
  author = {Enrico Brunetti and R. Neil Campbell and Jack Lovell and Dino A. Jaroszynski},
  title = {High-charge electron beams from a laser-wakefield accelerator driven by a CO2 laser},
  journal = {Sci Rep},
}

@article{WJPRAB2023,
  title = {High quality beam produced by tightly focused laser driven wakefield accelerators},
  author = {Wang, Jia and Zeng, Ming and Li, Dazhang and Wang, Xiaoning and Gao, Jie},
  journal = {Phys. Rev. Accel. Beams},
  volume = {26},
  issue = {9},
  pages = {091303},
  numpages = {8},
  year = {2023},
  month = {Sep},
  publisher = {American Physical Society},
  doi = {10.1103/PhysRevAccelBeams.26.091303},
  url = {https://link.aps.org/doi/10.1103/PhysRevAccelBeams.26.091303}
}

@article{IrshadPRR2023,
  title = {Multi-objective and multi-fidelity Bayesian optimization of laser-plasma acceleration},
  author = {Irshad, F. and Karsch, S. and D\"opp, A.},
  journal = {Phys. Rev. Res.},
  volume = {5},
  issue = {1},
  pages = {013063},
  numpages = {10},
  year = {2023},
  month = {Jan},
  publisher = {American Physical Society},
  doi = {10.1103/PhysRevResearch.5.013063},
  url = {https://link.aps.org/doi/10.1103/PhysRevResearch.5.013063}
}

@article{JalasPRAB2023,
  title = {Tuning curves for a laser-plasma accelerator},
  author = {Jalas, S. and Kirchen, M. and Braun, C. and Eichner, T. and Gonzalez, J. B. and H\"ubner, L. and H\"ulsenbusch, T. and Messner, P. and Palmer, G. and Schnepp, M. and Werle, C. and Winkler, P. and Leemans, W. P. and Maier, A. R.},
  journal = {Phys. Rev. Accel. Beams},
  volume = {26},
  issue = {7},
  pages = {071302},
  numpages = {6},
  year = {2023},
  month = {Jul},
  publisher = {American Physical Society},
  doi = {10.1103/PhysRevAccelBeams.26.071302},
  url = {https://link.aps.org/doi/10.1103/PhysRevAccelBeams.26.071302}
}

@article{FerranPRAB2023,
  title = {Bayesian optimization of laser-plasma accelerators assisted by reduced physical models},
  author = {Ferran Pousa, A. and Jalas, S. and Kirchen, M. and Martinez de la Ossa, A. and Th\'evenet, M. and Hudson, S. and Larson, J. and Huebl, A. and Vay, J.-L. and Lehe, R.},
  journal = {Phys. Rev. Accel. Beams},
  volume = {26},
  issue = {8},
  pages = {084601},
  numpages = {9},
  year = {2023},
  month = {Aug},
  publisher = {American Physical Society},
  doi = {10.1103/PhysRevAccelBeams.26.084601},
  url = {https://link.aps.org/doi/10.1103/PhysRevAccelBeams.26.084601}
}

@article{ZengPOP2024,
    author = {Zeng, Ming},
    title = "{Simulation observation of high effectiveness laser plasma wakefield accelerator using plasma telescope configuration}",
    journal = {Physics of Plasmas},
    volume = {31},
    number = {8},
    pages = {080702},
    year = {2024},
    month = {08},
    issn = {1070-664X},
    doi = {10.1063/5.0223051},
    url = {https://doi.org/10.1063/5.0223051}
}

@article{ShuangRe2024,
author = {Shuang Liu  and Fei Li  and Shiyu Zhou  and Jianfei Hua  and Warren B. Mori  and Chan Joshi  and Wei Lu },
title = {A Scalable, High-Efficiency, Low-Energy-Spread Laser Wakefield Accelerator Using a Tri-Plateau Plasma Channel},
journal = {Research},
volume = {7},
number = {},
pages = {0396},
year = {2024},
doi = {10.34133/research.0396},
URL = {https://spj.science.org/doi/abs/10.34133/research.0396}
}

\end{document}